# Frequency-comb-steered ultrawideband quasi-true-time-delay beamformer for integrated sensing and communication

Mian Wang, Wenxin Zhang, Zeyu Ren, Shangyuan Li, Xiaoping Zheng, and Xiaoxiao Xue[*]

*Department of Electronic Engineering, Beijing National Research Center for Information Science and Technology, Tsinghua University, Beijing 100084, China*
*[*xuexx@tsinghua.edu.cn](mailto:xuexx@tsinghua.edu.cn)*

**Abstract**

Phased array antennas (PAAs) possessing broadband beamforming capabilities are crucial for advanced radar and wireless communication systems. Nevertheless, traditional phase-shifter-based PAA beamformers frequently encounter the beam-squint issue, which substantially restricts their instantaneous bandwidth. Photonic true-time-delay (TTD) beamformers have the potential to overcome this challenge, offering ultrawide bandwidth and immunity to electromagnetic interference. However, their practical application is impeded by the high complexity, which typically involves a vast array of optical switches and delay lines. Here, we introduce a novel frequency-comb-steered photonic quasi-TTD beamformer that eliminates the need for delay lines by leveraging the concepts of frequency-diverse arrays and photonic microwave mixing arrays. This beamformer enables squint-free beamforming of ultrawideband linear frequency modulation waveforms, which is essential for high-resolution radar applications. It ensures seamless and continuous beam steering, effectively delivering infinite spatial resolution. We present a prototype with an 8-element PAA, demonstrating an instantaneous bandwidth of 6 GHz across the entire Ku-band. Additionally, we explore the system's capabilities in integrated inverse synthetic aperture radar imaging and high-speed communication, achieving a high imaging resolution of 2.6 cm × 3.0 cm and a transmission rate of 3 Gbps. Compared to conventional delay-line-based beamformers, our new concept markedly reduces hardware complexity and enhances scalability, positioning it as a potent enabler for future integrated sensing and communication applications.

**Introduction**

Phased array antennas (PAAs) have fundamentally transformed the way people manipulate electromagnetic waves in free space and played important roles in radar and communication applications. By controlling the phase of each antenna element, PAAs can generate flexible radiation patterns and steer the beam toward any specific direction [1]. Traditional PAA beamformers employ electronic phase shifters in each channel [2]. However, the well-known beam-squint issue that smears frequency components in space restricts the system to narrowband operations. The true-time-delay (TTD) technology can eliminate the beam squint by introducing time delays to the radiated signals [3]. With the ability to provide ultra-large instantaneous bandwidth, TTD beamformers show great potential for future high-resolution radar sensing [4] and high-capacity communications [5]. However, the high-loss, long-wavelength-scale microwave transmission lines inherently lead to lossy electronic TTD beamformers. Alternatives that promise low loss, such as artificial transmission lines [5,6] and switched filters [7], frequently encounter challenges with inconsistent broadband delay and the complexities of achieving precise impedance matching.

Photonic TTD beamformers have emerged as compelling contenders in superseding their electronic counterparts, leveraging the exceptional intrinsic benefits of photonic delay lines. These advantages include low propagation loss, a broad and flat delay spectrum, and robust immunity to electromagnetic interference [8-10]. One classic design of the photonic TTD beamformer is based on the programmable binary optical delay line which utilizes a series of optical switches and waveguides to alter the optical path length [11-13]. Photonic TTD beamformers built of optical fibers have long been investigated but the system is bulky and not suitable for large-scale PAAs [14]. Recent researches have increasingly focused on demonstrating integrated photonic TTD beamformers. Novel delay structures have been proposed for enhanced delay range and fine delay control, such as multimode waveguides [12], tunable Mach Zehnder interferometers [15,16], and slow-light devices [17,18]. Despite significant research endeavors, it is still particularly challenging for photonic TTD beamformers to support high-resolution and large-scale PAAs. As the number of antenna elements escalates alongside the need for greater resolution, the requisite number of optical switches and waveguides also surges, resulting in a marked increase in insertion loss and system complexity.

In this article, we introduce a novel frequency-comb-steered photonic quasi-TTD beamformer that eschews the need for delay lines, which substantially decreases complexity and enhances the scalability of the beamformer. The core concept is based on a photonically steered frequency diverse array, as depicted in Fig. 1(a). Traditionally, frequency-diverse arrays are designed to produce beampatterns that depend on both range and angle, which are limited to narrowband operations [19,20]. However, we demonstrate that by incorporating the well-known linear frequency modulation (LFM) and precisely controlling the frequency offsets between antenna channels, it is possible to achieve an elegant broadband quasi-TTD beamformer that boasts infinite angle resolution. Furthermore, we introduce a dual-comb heterodyne mixing scheme for massively parallel conversion of microwave frequencies [21-24]. Dual optical frequency combs, consisting of two sets of equally spaced discrete spectral lines, have emerged as pivotal bridges between the optical and microwave frequencies, with extensive applications in fields such as metrology and spectroscopy [25]. The dual comb mixing approach not only ensures an optimal linear frequency correlation across various antenna channels but also has the potential to support large-scale PAAs by exploiting a large number of comb lines. A proof-of-concept 8-element frequency-comb-steered quasi-TTD beamformer is presented, which achieves seamless squint-free beam steering across the entire Ku band (12-18 GHz). By leveraging its ultra-large instantaneous bandwidth of 6 GHz, we have successfully demonstrated joint high-resolution inverse synthetic aperture radar (ISAR) imaging and high-speed communication, achieving an imaging resolution of 2.6 cm × 3.0 cm and a communication data rate of 3 Gbps.

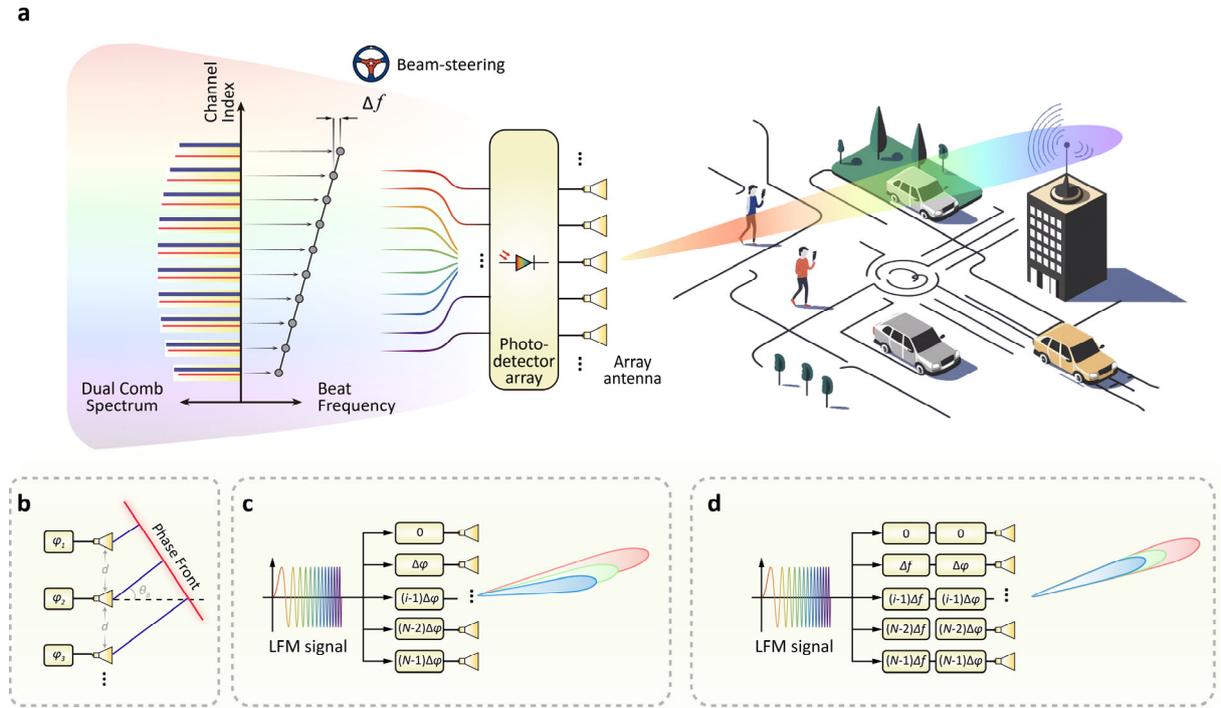

**Fig.1 | Concept of the frequency-comb-steered quasi-TTD beamformer. a**, Illustration of the beamformer. The microwave signals feeding the antenna array are generated by heterodyne mixing of dual optical frequency combs. Linear frequency increments with a frequency step $\Delta f$ are introduced across the antennas. Seamless and continuous beam steering can be achieved by changing $\Delta f$ and the phase increment of each antenna channel. **b**, Generalized principle of a linear antenna array. The electromagnetic waves interfere constructively at the phase front perpendicular to the beam direction. **c**, **d**, Comparison between a conventional phase-shifter-based array (**c**) and the proposed frequency-diverse quasi-TTD beamformer (**d**) for beamforming of LFM signals. The beam directions of different frequencies disperse for the conventional array. In contrast, all the beam directions are aligned to form a squint-free beam for the frequency-diverse quasi-TTD beamformer.

# Result

## Principle of frequency-comb-steered quasi-TTD beamformer

The generalized principle of a linear PAA is depicted in Fig. 1(b). The electromagnetic field intensity reaches its peak at the phase front with a distance $R$ and an angle $\theta_B$, provided that the following condition for constructive interference is met

$$\varphi_1(t - R/c) = \varphi_2(t - R/c - \tau_B) = ... = \varphi_N[t - R/c - (N-1)\tau_B] \tag{1}$$

Here, $\varphi_i(t)$ with $i = 1, 2, ..., N$ represents the instantaneous phase of the field from the $i$-th antenna; $c$ is the speed of light; $\tau_B$ is the differential time delay between adjacent antennas, given by $\tau_B = d \sin \theta_B / c$ where $d$ is the antenna spacing. For beamforming of LFM signals with traditional phase-shifter-based PAAs, the instantaneous phase of each antenna element can be written as

$$\varphi_i(t) = 2\pi \left( f_0 t + \kappa t^2 / 2 \right) + (i-1)\Delta\varphi \tag{2}$$

where $f_0$ is the instantaneous frequency at $t = 0$; $\kappa$ is the chirp rate; $\Delta\varphi$ is the differential phase shift between adjacent antennas. The beam direction according to Eq. (1) is approximately given by

$$\theta_B(R, t) \approx \sin^{-1} \left\{ \frac{\Delta\varphi}{2\pi \left[ f_0 + \kappa(t - R/c) \right]} \frac{c}{d} \right\} \tag{3}$$

Obviously, the beam direction is range- and time-dependent (and therefore also dependent on the instantaneous frequency; see Fig. 1(c)). However, if a linear frequency offset is simultaneously introduced across the antenna array, the instantaneous phases are given by

$$\varphi_i(t) = 2\pi \left( f_0 t + \kappa t^2 / 2 \right) + (i-1)(2\pi \Delta f t + \Delta\varphi) \tag{4}$$

where $\Delta f$ is the frequency increment. The instantaneous frequency at $t = 0$ for the $i$-th channel is thus $f_i = f_0 + (i-1)\Delta f$ and the additional phase shift for the $i$-th channel is $\Delta\varphi_i = (i-1)\Delta\varphi$. It can be verified that the beam direction will be time-invariant if the following conditions are satisfied

$$\Delta f = \kappa \tau_B \text{ and } \Delta\varphi = 2\pi f_0 \tau_B \tag{5}$$

The beam direction is given by

$$\theta_B = \sin^{-1} \frac{\Delta f c}{\kappa d} \tag{6}$$

Beam scanning can be achieved simply by changing $\Delta f$ and $\Delta \varphi$. It should be noted that the LFM signal may have a very large instantaneous bandwidth. Nevertheless, similar to a delay-line-based TTD beamformer, there is no beam squint at all instantaneous frequencies in this configuration, as illustrated in Fig. 1(d) (see Supplementary Note I for the derivation of far-field beampatterns).

Constructing a large-scale frequency-diverse array traditionally demands an extensive array of electronic mixers and synchronized local oscillators (LOs), each operating at distinct frequencies. This approach is not only complex but also consumes significant power. To surmount these challenges, we employ a massively parallel photonic microwave mixer array that leverages dual optical frequency combs. As depicted in Fig. 1(a), the dual-comb configuration comprises two optical frequency combs with slightly different repetition rates. One comb is modulated by the microwave signal, while the other acts as the LO comb. Upon combining the two combs, the comb line pairs are demultiplexed and detected by an array of photodetectors (PDs), generating a series of microwave signals with linearly increasing frequencies and phase offsets. For the $i$-th channel, the instantaneous microwave phase compared to the first channel is given by

$$\Delta \varphi_i = (i-1)(2\pi \Delta f t + \Delta \varphi) \qquad (7)$$

where $\Delta f$ is the difference in comb repetition rates and $\Delta \varphi$ is the initial phase offset at $t=0$. In addition to significantly reducing hardware complexity, the dual-comb-based photonic microwave mixer array boasts another notable advantage – the frequency conversion is inherently synchronized across channels. This synchronization is a natural consequence of the mode-locking properties inherent to optical frequency combs.

**Seamless squint-free beamforming of ultra-wideband LFM signals**

The experimental setup of an 8-element frequency-comb-steered quasi-TTD beamformer is shown in Fig. 2(a). We employ electro-optic (EO) combs for the dual-comb heterodyne mixing scheme. Among the various techniques for frequency comb generation [26,27], EO combs stand out for their exceptionally flat spectral profile and the capability to adjust the repetition rate with great flexibility, making them ideal for frequency-diverse array applications [28,29]. A continuous-wave (CW) laser serves as the seed light which is split into two distinct paths – the signal path and the LO path – to create dual EO combs. The combs are produced using an intensity modulator (IM) followed by a phase modulator (PM). In the signal path, the modulators are driven by a single-tone microwave at $f_{RF1} = 31 \text{ GHz}$. Subsequently, the generated signal comb is modulated by a 2-8 GHz LFM microwave signal through carrier-suppressed single-sideband (CS-SSB) modulation. In the LO path, the CW light is frequency-shifted by $f_{shift} = 10$ GHz through CS-SSB modulation. Then the modulators are driven by a single-tone microwave at $f_{RF2} = (31 + \Delta f)$ GHz to generate the LO comb. The signal and LO combs are merged, demultiplexed, and detected by a PD array. The frequency of the generated LFM signals is upconverted to the 12-18 GHz range, covering the entire Ku band. These signals are subsequently directed to an 8-element Vivaldi antenna array with an element spacing of 23 mm, as shown in the inset of Fig. 2(a). It is worth highlighting that our dual-comb scheme seamlessly integrates photonic microwave frequency up-conversion with beamforming, thereby substantially diminishing the reliance on high-frequency electronic components. Moreover, the system can be easily upgraded to the millimeter wave and terahertz bands by employing readily available larger-bandwidth modulators and PDs.

Figure 2(b) shows the optical spectrum of the combined dual combs, as measured at the input of the wavelength demultiplexer. Over 16 pairs of comb lines are generated within a 3-dB bandwidth. The sharper teeth represent the LO comb, while the wider teeth represent the signal comb modulated by the LFM signal. Eight pairs of comb lines highlighted in Fig. 2(b) are selected for beamforming. To illustrate the principle of frequency-comb-steered beamforming, the experimental results of beam steering at angles of −30°, −15°, 0°, +15°, and +30° are shown in Figs. 2(c)-(e), respectively. The pulse width of the transmitted 12-18 GHz LFM signal is 20 μs, corresponding to a chirp rate of 0.3 GHz/μs. To steer the beam direction, the relative frequencies and initial phase offsets of the two microwave tones that drive the dual combs are adjusted in accordance with Eq. (5). The resultant frequency and phase offsets for each channel, compared to the 4[th] channel, are depicted in Figs. 2(c) and 2(d). The far-field waveforms of the antenna array are captured with a high-gain Vivaldi horn antenna positioned at various angles and are subsequently recorded by a real-time oscilloscope. The beam patterns at different frequencies are then obtained by performing Fast Fourier Transform (FFT) on the received signals. The results at 15 GHz are shown in Fig. 2(e), clearly illustrating the system's capability for agile beam steering. The reduction in beam power at angles that deviate from 0° can be ascribed to the non-omnidirectional radiation pattern of the antenna array elements (see Supplementary Fig. S3). It is worth noting that the beam direction can be easily manipulated by simply altering the frequency and phase of the microwave tone that drives the LO comb (i.e., $f_{RF2}$). This configuration presents a significant simplification compared to the conventional delay-line-based TTD beamformer, which requires individual control of the delay for each channel.

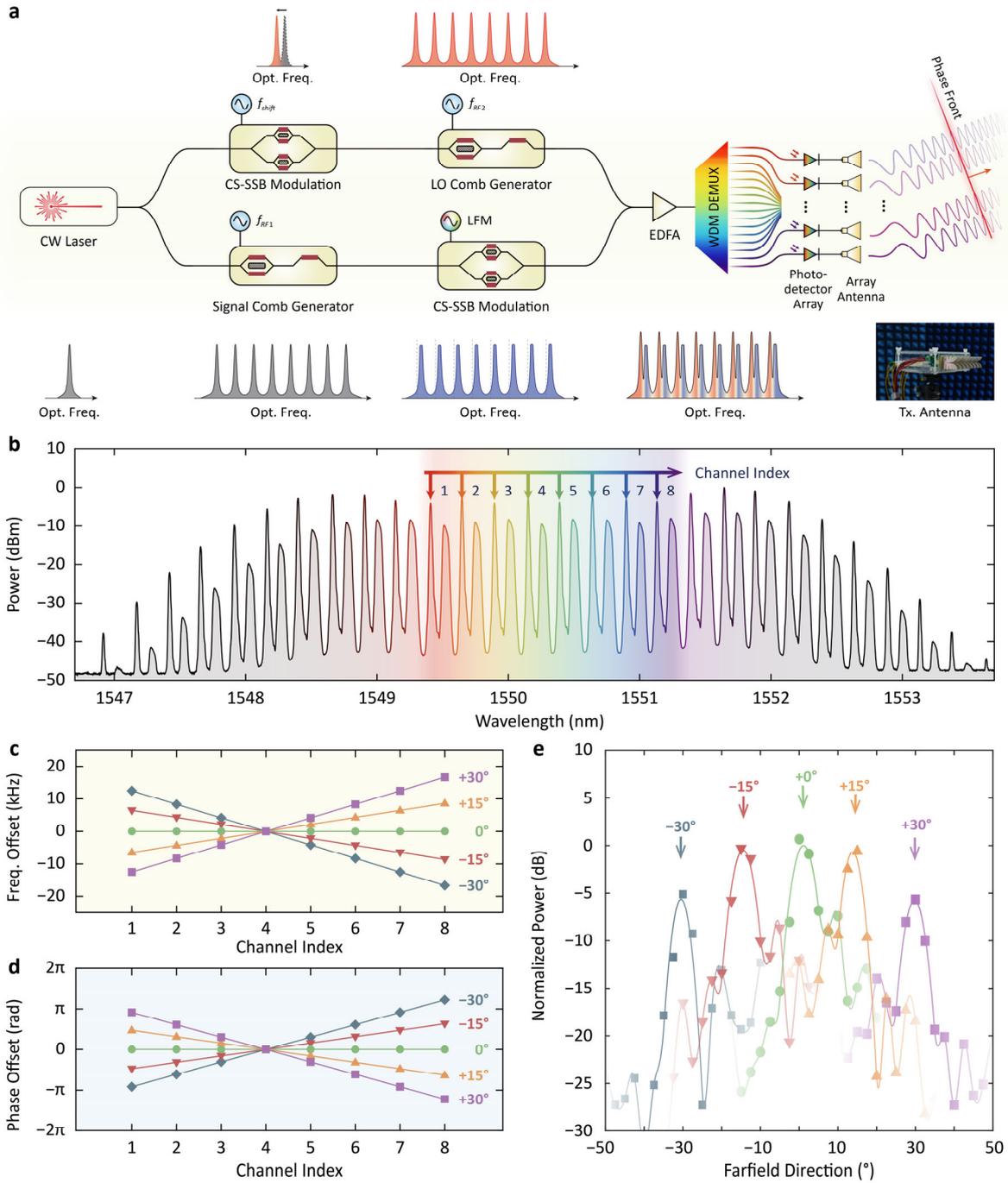

**Fig.2 | Demonstration of frequency-comb-steered quasi-TTD beamformer. a**, Experimental setup. Dual frequency combs (i.e., signal comb and LO comb) are generated with cascaded modulators. The signal comb is modulated by an intermediate-frequency LFM signal, while the LO comb is frequency-shifted through carrier-suppressed single-sideband (CS-SSB) modulation. The dual combs are then amplified, demultiplexed, and heterodyned to generate microwave signals that feed an 8-element Vivaldi antenna array. The inset shows the antenna's picture. EDFA, erbium-doped fiber amplifier. WDM DEMUX, wavelength division multiplexing demultiplexer. **b**, Spectrum of the dual combs. The highlighted eight channels are used for beamforming. **c, d,** Frequency and phase offset of the transmitted LFM signals in each channel, compared to the 4th channel. **e**, Radiation patterns at the instantaneous frequency of 15 GHz when the beam direction is −30°, −15°, 0°, +15°, and +30° respectively. The main lobes are highlighted. Markers: measured; solid lines: fitted.

Figure 3 illustrates the beamforming performance across a wide bandwidth. The beamforming results for the directions of +30°, 0°, and −30° are detailed in Figs. 3(a), 3(b), and 3(c), respectively. Each subplot displays the transient beam patterns at three distinct frequencies: 12 GHz, 15 GHz, and 18 GHz. The alignment of the main lobes indicates the absence of beam-squint issues. Furthermore, the space-frequency contour plots presented in Figs. 3(d), 3(e), and 3(f) demonstrate that the main lobe direction is maintained consistently throughout the entire LFM bandwidth, with a remarkable agreement between experimental and simulation results. It is noteworthy that grating lobes are present in our current demonstration. However, the spatial distribution of these grating lobes is highly dispersed, leading to a greatly reduced grating lobe intensity for any specific direction angle. The impact of grating lobes may further be diminished in the future by reducing the antenna spacing to less than half the shortest wavelength of the LFM signal. Figure 3(g) depicts the 12-18 GHz LFM signals received at −30° and +30°. The waveform envelope remains continuous throughout the full 20-μs pulse duration, without obvious fading, thereby further demonstrating the squint-free beamforming characteristic. The minor variation of the envelope is attributed to the broadband frequency response of the transmitting and receiving antennas (see Supplementary Fig. S2).

An additional noteworthy advantage of the dual-comb-based beamformer is its ability to continuously scan the beam direction by simply adjusting the frequency and phase offset of the microwave source that drives one of the two combs. Figure 3(h) shows the result of beam steering with an azimuth step of 1°. The microwave frequency shift needed to adjust the beam direction by 1° is approximately 500 Hz. Finer steering steps, such as 40 Hz frequency adjustment for a 0.1° beam steering, can also be easily achieved by modifying the microwave source frequency. The maximum frequency shift required for beam steering across a range of −30° to +30° is around 23 kHz — well within the capabilities of current electronic systems. This straightforward tunability, coupled with ultra-high resolution, greatly reduces the complexity of the beam control system. In contrast, a traditional 8-element TTD beamformer relying on binary delay lines would require at least 9-bit resolution to steer the beam over the same −30° to +30° range with a 0.1° step [8]. This would entail at least 72 optical switches and waveguides that require high-precision fabrication and control, posing substantial challenges.

**Integrated radar imaging and communication**

Integrated sensing and communication (ISAC) technology [30-33] is garnering increasing interest due to its significant potential in applications such as unmanned vehicles, smart homes, and the Internet of Things [33-35]. In this section, we present the capabilities of our frequency-comb-steered ultrawideband beamformer for the dual functions of high-resolution ISAR imaging and high-speed communication. The dual-function waveform we utilized for ISAC is constant envelope (CE) linear frequency modulation with orthogonal frequency division multiplexing (LFM-OFDM) [36,37]. This waveform leverages the pulse compression benefits of broadband LFM signals and the multipath resistance of OFDM. By embedding real-valued OFDM symbols onto the phase of the LFM carrier, the CE-LFM-OFDM modulation scheme generates waveforms with a constant envelope, effectively preventing nonlinear distortions in power amplifiers [38,39].

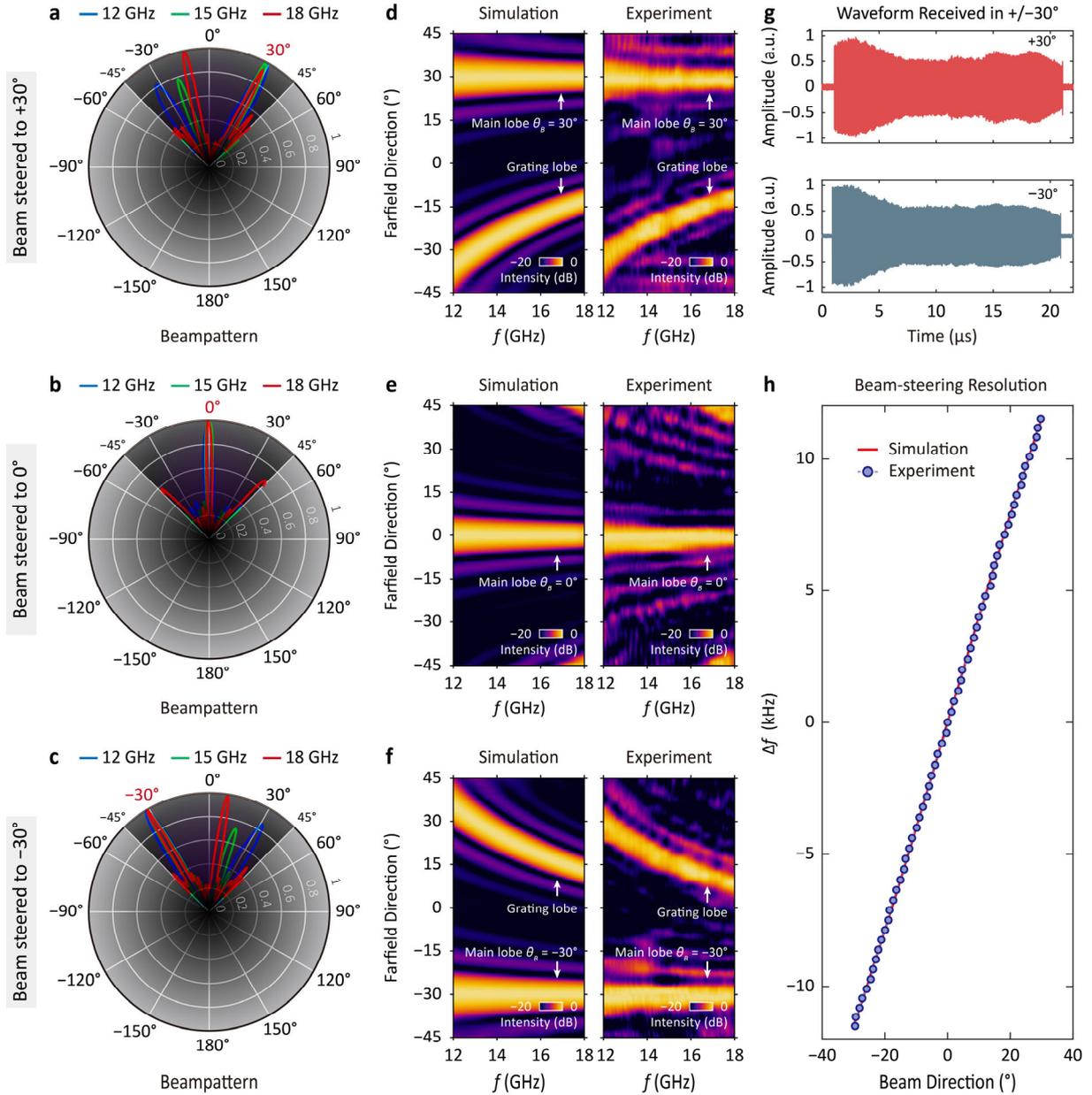

**Fig.3 | Experimental results of broadband beamforming and continuous beam steering. a**, **b**, **c**, Polar plots of radiation patterns at 12 GHz, 15 GHz, and 18 GHz when the beam is steered to +30°, 0°, and −30° respectively. **d**, **e**, **f**, Space-frequency contour that depicts consistent beam direction throughout 12-18 GHz. The experimental results (right) show remarkable agreement with the simulations (left). Notably, the large array spacing is responsible for the grating lobes that exist in (a)~(f), which can be eliminated by utilizing antenna arrays with element spacing smaller than half the minimum operating wavelength. **g**, Time domain waveforms detected with a receiving antenna placed in the beam direction when the beam is steered to −30° and +30°, respectively. **h**, Demonstration of seamless beam steering. The beam direction is scanned with an azimuth step of 1° in experiments.

The experimental configuration is depicted in Fig. 4(a). The CE-LFM-OFDM waveform, operating in the 12-18 GHz LFM band with a symbol rate of 500 Mbaud and 64-QAM modulation (corresponding to a bit rate of 3 Gbps), is transmitted by the frequency-comb-steered PAA. An imitated radar target, comprising a four-point scatterer assembly mounted on a rotating pedestal, is positioned 1.5 m away from the PAA. For imaging purposes, a receiving horn antenna is positioned next to the PAA transmitter, while another horn antenna is placed near the target for communication. The transmitted and received CE-LFM-OFDM waveforms are shown in Figs. 4(b) and 4(c), with the received waveform captured by the communication receiver. For imaging, a photonic de-chirp scheme is employed to capture the radar echoes. It performs optical mixing of the echoes with transmitted radar signal, producing a demodulated signal with a frequency of MHz level [4,40]. The resulting target image, captured when the beam is directed at 0° and −30° respectively and the target is aligned with the beam, is shown in Figs. 4(d) and 4(e). Four distinct scatterers are clearly visible with a spatial resolution of $2.6 \text{ cm} \times 3.0 \text{ cm}$. Concurrently, the demodulated constellation of the communication data is depicted. The error vector magnitudes (EVMs) are 5.2% and 7.0% for the beam angles of 0° and −30° respectively, complying with the maximum permissible EVM of 8.0% as stipulated by the 3rd Generation Partnership Project (3GPP) for 64-QAM modulation. As anticipated, both ISAR imaging and symbol recovery are unsuccessful when the target and the communication receiver are positioned outside the beam's coverage area.

The variation of EVM in relation to the radar bandwidth and symbol rate are shown in Figs. 4(f) and 4(g), respectively. All tests were conducted with the carrier LFM centered at 15 GHz. As can be observed, the LFM bandwidth for radar imaging has a negligible impact on the communication EVM at the tested data rates. For a fixed radar bandwidth, the EVM increases as the symbol rate rises, as shown in Fig. 4(g). This increase is primarily due to the reduced energy per bit at higher symbol rates, leading to a lower signal-to-noise ratio (SNR) per bit and a degradation in demodulation accuracy. Higher SNR and lower EVM are within reach by using microwave amplifiers with higher gain at the transmitter.

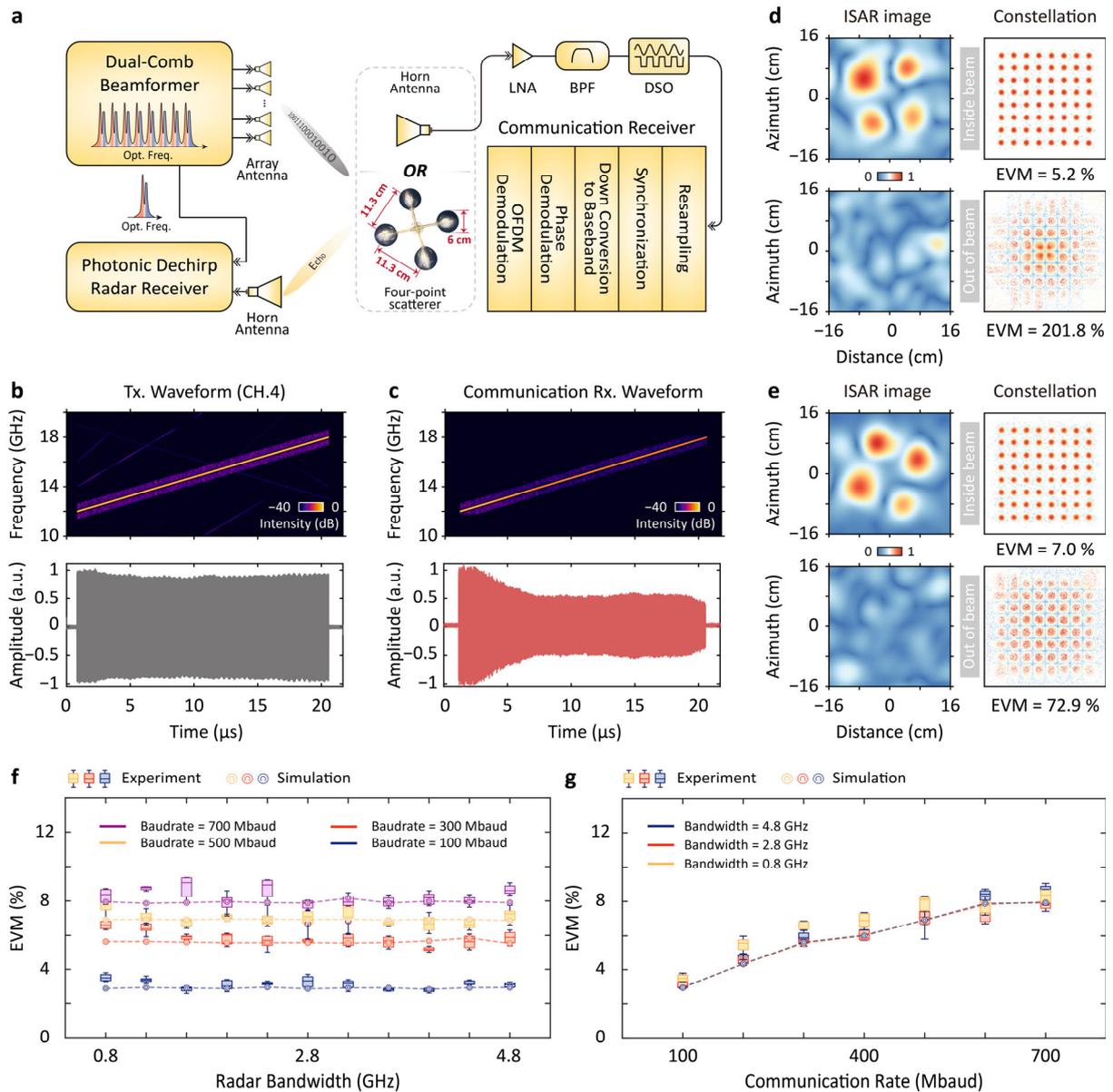

**Fig.4 | Integrated radar imaging and communication with the frequency-comb-steered quasi-TTD beamformer**. **a**, Experimental setup. The radar receiver is co-located with the PAA transmitter, while the communication receiver and the 4-point scatterer are positioned 1.5 meters away. The communication receiver collects signals by a Vivaldi horn antenna. After being amplified and filtered, the received signals are digitalized by a DSO, followed by demodulation of communication symbols in a computer. LNA, low noise amplifier; BPF, bandpass filter; DSO, digital storage oscilloscope. **b**, **c**, Spectrograms and waveforms of the 12-18 GHz, 500 Mbaud, 64-QAM CE-LFM-OFDM signal. The transmitted (**b**) and received (**c**) signals were acquired at the input of the 4th antenna and at the output of BPF in the communication receiver, respectively. **d**, **e**, Results of ISAR imaging and communication EVM when the beam direction is 0° (**d**) and −30° (**e**) respectively. Instances where the target and communication receiver are either aligned with the antenna beam (upper) or misaligned (lower) are depicted. **f**, Communication EVM versus radar bandwidth with different communication rates. **g**, Communication EVM versus communication rate with different radar bandwidths.

# Discussion

In conclusion, we have demonstrated a frequency-comb-steered delay-line-free quasi-TTD beamformer capable of broadband beamforming with infinite steering resolution. The proposed system was validated through successful beamforming at various angles with an 8-element antenna array transmitting 12-18 GHz LFM signals. Furthermore, we extended this system to integrated ISAR imaging and communications, achieving a high imaging resolution of 2.6 cm×3.0 cm and a data rate of 3 Gbps with 64-QAM, 500 Mbaud CE-LFM-OFDM signals. These results demonstrate the system's versatility for emerging multi-functional applications that demand wideband beamformers.

Future work will focus on miniaturization and integration of the proposed beamformer. Advances in integrated EO comb generation have promised tens to hundreds of comb teeth that could enable massive-scale arrays [41,42]. Recent advances in integrated photonic circuits may facilitate the integration of laser sources [43], modulators [44], amplifiers [45], and PDs [46] onto a single chip, paving the way for compact, power-efficient implementations suitable for mobile or space-constrained platforms [47,48]. With these improvements, the proposed frequency-comb-steered quasi-TTD beamformer will provide an efficient and scalable solution for broadband applications, meeting the growing demand for high-frequency, large-scale arrays in next-generation radar and communication systems.

# Methods

### Dual-comb generation

A continuous-wave laser operating at 1550.2 nm is utilized as the seed light source. The signal and LO combs are produced using two distinct cascaded configurations of intensity modulators and phase modulators. Each comb generator is driven by a separate microwave source, with synchronization achieved through a 10 MHz clock signal. Within the signal path, the comb repetition rate is set at 31 GHz. A LFM signal, spanning 2-8 GHz with a 20-μs pulse duration, is generated by an arbitrary waveform generator and subsequently modulated onto the signal comb using a dual-parallel Mach-Zehnder modulator biased for carrier-suppressed single-sideband (CS-SSB) modulation. In the LO path, the seed light frequency is upshifted by 10 GHz through CS-SSB modulation before the generation of the LO comb. The LFM signals, resulting from the heterodyning of the signal comb with the LO comb, are then upconverted by 10 GHz, spanning a frequency range from 12 to 18 GHz. The LO comb's repetition rate is minutely detuned from that of the signal comb by a few kilohertz, as calculated using Eq. (5).

**ISAR imaging**

The radar ISAR imaging is carried out by transmitting a sequence of pulses of CE-LFM-OFDM signal. The emulated target is a four-scatterer assembly (containing four metallic balls) mounted on a rotating pedestal with a rotational speed of 2.4 rev/s. The target is positioned 1.5 m away from the beamformer. A photonic radar de-chirp receiver is used to capture the echoes [4,42]. In the receiver, the transmitted radar signal is optically mixed with the received target echoes by using an optical intensity modulator to generate a de-chirped intermediate frequency (IF) signal in the MHz frequency band. The IF signal is then amplified, filtered, and recorded by an oscilloscope for subsequent processing. To reconstruct the ISAR image, a two-dimensional Fourier transform approach is used [49]. First, an in-pulse FFT is performed to obtain the target's range profile. Next, an inter-pulse FFT is applied to extract azimuth information by analyzing the Doppler frequency shifts across successive pulses. Each frame of the target's two-dimensional ISAR image is generated by processing 1200 pulses.

**CE-LFM-OFDM modulation and demodulation**

To generate the CE-LFM-OFDM signal, inverse discrete Fourier transform (IDFT) is first used to generate the time-domain samples of each OFDM frame carrying $N_{QAM}$ data symbols. A conjugate symmetric data vector is constructed as the input to the IDFT to ensure real-valued OFDM [37]. The size of IDFT is given by $N_{DFT} = 2N_{QAM}+N_{ZP}+2 = 1024$, where $N_{ZP}$ is the number of zero-padding points added for oversampling. A cyclic prefix is inserted into each OFDM symbol to mitigate inter-symbol interference. The IDFT output is then modulated onto the phase of the LFM signal, with a phase modulation index of 0.5. The memory phase is introduced in phase modulation to achieve smooth phase transitions and minimize spectral leakage [50]. At this point, the CE-LFM-OFDM waveform is complete and ready for transmission. In the receiver, the demodulation process is inverse to that of modulation. First, the received CE-LFM-OFDM signals are down-converted to the baseband, followed by phase demodulation to extract the real-valued OFDM symbols with cyclic prefixes. The cyclic prefixes are then removed, and the discrete Fourier transform (DFT) is performed to recover the QAM symbols carried by the subcarriers. Frequency-domain equalization is performed to correct the channel-induced distortions.